\documentclass[onecolumn,useAMS,usenatbib]{mn2e}
\usepackage{epsfig}
\usepackage{verbatim}
\usepackage{hyperref}
\usepackage{multicol}
\usepackage{amsmath,amssymb,comment,subfig}
\usepackage{framed,graphicx,multirow,array}
\usepackage[usenames,dvipsnames]{color}

\newcommand{\beeq}{\begin{equation}}
\newcommand{\eneq}{\end{equation}}
\newcommand{\bear}{\begin{eqnarray}}
\newcommand{\enar}{\end{eqnarray}}
\newcommand{\bef}{\begin{figure*}}
\newcommand{\enf}{\end{figure*}}

\title[Early-type galaxy alignments and CMB lensing]{Contamination of early-type galaxy alignments to galaxy lensing-CMB lensing cross-correlation}
\author[Chisari et al.]{Nora Elisa Chisari,$^1$\thanks{elisa.chisari@physics.ox.ac.uk} Joanna Dunkley,$^1$ Lance Miller$^1$ and Rupert Allison$^1$ \\
$^1$ Department of Physics, University of Oxford, Keble Road, Oxford,\\
OX1 3RH, United Kingdom.}

\date{In current form \today}

\pagerange{\pageref{firstpage}--\pageref{lastpage}} \pubyear{2014}

\begin{document}

\label{firstpage}

\maketitle

\begin{abstract}
Galaxy shapes are subject to distortions due to the tidal field of the Universe. The cross-correlation of galaxy lensing with the lensing of the Cosmic Microwave Background (CMB) cannot easily be separated from the cross-correlation of galaxy intrinsic shapes with CMB lensing. Previous work suggested that the intrinsic alignment contamination can be $15\%$ of this cross-spectrum for the CFHT Stripe 82 (CS82) and Atacama Cosmology Telescope surveys. Here we re-examine these estimates using up-to-date observational constraints of intrinsic alignments at a redshift more similar to that of CS82 galaxies. We find a $\approx$ $10\%$ contamination of the cross-spectrum from red galaxies, with $\approx$ $3\%$ uncertainty due to uncertainties in the redshift distribution of source galaxies and the modelling of the spectral energy distribution. Blue galaxies are consistent with being unaligned, but could contaminate the cross-spectrum by an additional $9.5\%$ within current $95\%$ confidence levels. While our fiducial estimate of alignment contamination is similar to previous work, our work suggests that the relevance of alignments for CMB lensing-galaxy lensing cross-correlation remains largely unconstrained. Little information is currently available about alignments at $z>1.2$. We consider the upper limiting case where all $z>1.2$ galaxies are aligned with the same strength as low redshift luminous red galaxies, finding as much as $\approx$ $60\%$ contamination.
\end{abstract}


\section{Introduction}

Large-scale tidal forces stretch galaxy ellipsoids \citep{Catelan01}, elongating their shapes in the direction of nearby overdensities. This stretching results in a net projected ellipticity that contaminates weak gravitational lensing, ``shear", measurements \citep{Hirata04,Hirata10}. While lensing by the intervening matter creates a tangential alignment of background source galaxies around the lens \citep{Kaiser92}, tides have the opposite effect, typically leading to a radial alignment of sources towards overdensities. Alignments across large separations were first measured by \cite{Brown02}, with a subsequent confirmation by \cite{Mandelbaum06}. This signal was shown to arise from the alignment of luminous red galaxies \citep{Hirata07,Joachimi11,Singh14}, while it has not been detected for late-type galaxies \cite[e.g.][]{WiggleZ,Heymans13}. For this population, it is expected that the tidal field of the Universe would perturb the orientation of the disk by applying a torque on the galaxy angular momentum, which results in a smaller degree of alignment than for red galaxies \citep{Catelan01,Hirata04}. For a review on intrinsic alignments see \cite{Troxel14b}.

Shear from gravitational lensing and from intrinsic alignments are hard to disentangle; the signals have opposite signs and the cross-spectrum of the shear and density field correlation is thus reduced by the contribution of alignments. Upcoming imaging surveys, such as {\it Euclid}\footnote{\url{http://sci.esa.int/euclid/}} and {\it LSST}\footnote{\url{http://www.lsst.org/lsst/}}, will attempt to constrain the equation of state of dark energy by measuring the shapes of billions of galaxies to determine the lensing due to the intervening matter distribution. Potential biases in these constraints from intrinsic alignment contamination are significant \citep{Kirk12} and several techniques have been proposed to minimize the impact of these alignments \citep{Joachimi10a,Joachimi10b,Zhang10}.

\cite{Hand13} made the first measurement of the cross-correlation of galaxy lensing with the lensing of the CMB. This measurement constrains structure formation at intermediate redshifts, and it is advantageous due to the fact that the systematics in the galaxy shape measurement and in the CMB fluctuations cancel out in cross-correlation. \cite{Hand13} initially found a discrepancy between the predicted amplitude of the cross-correlation signal ($A=1$) and the best fit amplitude: $A^{\rm WMAP}=0.74\pm0.23$ for WMAP$9$ cosmology \citep{Hinshaw13} and $A^{Planck}=0.61\pm0.19$ for {\it Planck} cosmological parameters \citep{Planck}. They suggested that this discrepancy could be explained by uncertainties in the photometric redshift distribution of the galaxy sample from the CS82 survey \citep{Shan14}. With a re-analysis of the data which improved on the normalization procedure for shear maps, these amplitudes derived in \cite{Hand13} have now been updated to $A^{\rm WMAP} = 0.92 \pm 0.22$ and $A^{Planck} = 0.78\pm 0.18$.

\cite{Troxel14a} and \cite{Hall14} suggested that the contamination of intrinsic alignments could explain part of the original discrepancy in the amplitude of the measured cross-correlation relative to expectations from WMAP$9$ and {\it Planck}. They estimated a contamination of $15\%$ from intrinsic alignments to the cross-correlation of galaxy shapes and CMB lensing. However, these estimates were based on assuming that all galaxies in the CS82 survey are subject to intrinsic alignments with an amplitude given by the SuperCOSMOS survey \citep{Brown02,Hirata04,Bridle07}. This amplitude was measured for all galaxies in SuperCOSMOS down to a magnitude of $b_J\leq 20.5$ over an area of $10,000$ deg$^2$. State-of-the-art observations of intrinsic shape correlations suggest that only early-type galaxies are subject to alignments within current observational uncertainties. For example, \cite{Heymans13} measured intrinsic alignments in the CFHTLenS survey, classifying them as early and late-type using their best fit spectral template. This results in $80\%$ of the galaxies in their catalogue being late-type. For these galaxies, \cite{Heymans13} find no indication of intrinsic shape correlations, while these correlations are detected at the $3\sigma$ level for early-types. \cite{WiggleZ} also found blue galaxy alignments to be consistent with null for an intermediate redshift sample with imaging from the {\it Sloan Digital Sky Survey} \citep[][SDSS]{Eisenstein01} and spectroscopic redshifts from the WiggleZ Dark Energy Survey \citep{Parkinson12}.  We can expect that similarly, in the CS82 survey, the majority of the galaxies will be late-types \citep{Strateva01}. Moreover, the intrinsic alignment amplitude is observed to depend on galaxy luminosity \citep{Hirata07}, and it has now been measured to much higher redshift than for SuperCOSMOS \citep{Joachimi11,Singh14}. Hence, a more accurate model for the contamination from alignments to the CMB lensing-galaxy lensing cross-correlation should include the modelling of the fraction of early-type galaxies in CS82 and their intrinsic alignment strength as a function of luminosity and redshift.

 Recently, \citet{Liu15} measured the cross-correlation of CMB lensing and galaxy-lensing using from the combination of {\it Planck} and CFHTLenS\footnote{\url{http://www.cfhtlens.org}}. They found a larger discrepancy between the expected amplitude of CMB lensing-galaxy lensing cross-correlation and their measurements than found in \citet{Hand13}. This is quantified by $A^{\rm WMAP}_{\rm LH15}=0.52\pm0.26$ and $A^{\rm Planck}_{\rm LH15}=0.44\pm0.22$, assuming WMAP9 and {\it Planck} cosmologies, respectively. These authors suggest that possible reasons for the devitation of $A$ from $1$ are: intrinsic alignments ($15\%$ relying on the \citealt{Troxel14a} and \citealt{Hall14} estimates), masking of thermal Sunyaev-Zeld'ovich clusters in the CMB lensing reconstruction ($10\%$), photometric redshift uncertainties ($10\%$) and unknown multiplicative biases in the CFHTLenS shear calibration.

In this work, we use a model for the early-type galaxy population and their alignments in the redshift range spanned by the CS82 survey and we give a new estimate of the contamination from alignments to the galaxy lensing-CMB lensing cross-correlation to the \citet{Hand13} results. Section \ref{sec:theory} introduces the intrinsic alignment model. Section \ref{sec:redgal} describes the modeling of the red galaxy population based on the approach proposed by \cite{Joachimi11}. In Section \ref{sec:results}, we estimate the contamination of intrinsic alignments to the cross-correlation of galaxy shapes with CMB lensing. Finally, we list some of the limitations for future surveys in Section \ref{sec:forecast}. Throughout this paper, we assume the following \cite{Ade:2015} fiducial cosmology: $\Omega_{\rm b}h^2=0.022$, $\Omega_{\rm CDM}h^2=0.12$, $h=0.67$, $\Omega_K=0$, $A_s=2.2\times10^{-9}$, $n_s=0.9645$, $k_p=0.05$ Mpc$^{-1}$ and we define $\Omega_M=\Omega_b+\Omega_{\rm CDM}$.

We will generally refer to early-type galaxies as ``red'' galaxies, although the model presented in Section \ref{sec:redgal} considers the possibility that these galaxies maintain a small constant star formation rate. It is known that a small fraction ($\sim 6\%$) of low-redshift early-type galaxies display blue colors \citep{Schawinski09}. While it is possible that blue early-types are also subject to alignments, the overall alignment strength of blue galaxies is consistent with null up to $z=1.3$ \citep{WiggleZ,Heymans13} and so we do not consider them as contaminants in general. We discuss the details of this assumption in Section \ref{sec:results}.

\section{Intrinsic alignment model and weak gravitational lensing}
\label{sec:theory}

The intrinsic alignments of ellipsoidal galaxies are manifested in a change of the ellipticity of the galaxy caused by the tidal field of the Universe. Following \cite{Catelan01} and \cite{Hirata04}, the intrinsic shape-matter overdensity cross-power spectrum can be modeled by

\begin{equation}
P_{\delta I}(k,z) = - \frac{A_I(\langle L_r \rangle) C_1 \rho_{\rm crit} \Omega_m}{D(z)} P_{\delta}(k,z),
\label{eq:iaps}
\end{equation}

\noindent where $D(z)$ is the growth function of matter fluctuations normalized to unity today. This model arises from assuming that galaxy intrinsic ellipticities are linearly related to the tidal field with an unknown proportionality constant, $A_I$. This constant quantifies the degree of alignment between red galaxies and the tidal field as a function of the mean $r$-band luminosity of the sample. $C_1\rho_{\rm crit}=0.0134$ is a normalization fixed by convention to the results of the SuperCOSMOS intrinsic alignment measurements \citep{Brown02,Hirata04,Bridle07}. Eq. (\ref{eq:iaps}) also relies on the assumption that galaxies adopt preferential elongations at the time of formation and they preserve that information throughout the assembly of the large-scale structure. We use the non-linear matter power spectrum $P_\delta(k,z)$ from CAMB \citep{CAMB}, which uses the HALOFIT correction \citep{HALOFIT}. While this procedure does not fully capture the non-linear evolution of alignments, it has been shown to reproduce alignments on small scales with reasonable accuracy and it is widely used in the literature \citep{Hirata07,Bridle07,Blazek11,Chisari13}.

\citet{Joachimi11} measured the strength of alignment from a sample of Luminous Red Galaxies (LRGs) in the SDSS as a function of redshift and $r$-band luminosity. They found that the non-linear alignment model is a good fit to the data, with a normalization constant of

\begin{equation}
A_I(L_r) = (5.76^{+0.60}_{-0.62}) \left( \frac{ L_r }{L_0}\right)^{1.13^{+0.25}_{-0.2}},
\label{eq:AIJ11}
\end{equation}
where $L_0$ is a pivot luminosity corresponding to an absolute magnitude of $M_0(z)=-22-1.2z-5\log_{10}h$ where the redshift evolution of $M_0$ is accounted for in the $1.2z$ term \citep{Faber07} and we correct for our use of $h=0.67$ rather than $h=1$ as in \cite{Joachimi11}. We do not consider redshift evolution of the intrinsic alignment amplitude beyond that of the ``primordial'' alignment model of \cite{Catelan01}, as \cite{Joachimi11} find it to be consistent with that model.
More recently \cite{Singh14} measured the strength of alignment from the LOWZ sample of red galaxies in the {\it Sloan Digital Sky Survey III} \citep{Dawson13}, including its dependence on redshift and luminosity. The LOWZ galaxies are selected to extend the LRG population to fainter magnitudes. \citet{Singh14} found a redshift evolution consistent with primordial alignments, and an $r$-band luminosity dependence in agreement with the results of \citet{Joachimi11} for brighter galaxies. They also found evidence of an excess of power in the intrinsic alignments of LOWZ galaxies on scale $<1$ Mpc$/h$ with respect to the non-linear alignment model, which they modelled using a halo model for alignments \citep{Schneider10}. However, CMB lensing has little sensitivity to such scales, and we thus do not consider them in this work.

The convergence field inferred from galaxy shapes has two components, $\kappa_S = \kappa_G+\kappa_I$. One is due to weak gravitational lensing, $\kappa_G$ ($G$, in short), while the other one is due to intrinsic shapes, $\kappa_I$ (later referred to as $I$). All galaxies contribute to $\kappa_G$, while only early-type galaxies are thought to contribute to $\kappa_I$.

The cross-correlation of the CMB lensing convergence field and the convergence from galaxy shapes has thus two components, as described in \cite{Hall14,Troxel14a}. The intrinsic alignment component can be modeled by weighting Eq. (\ref{eq:iaps}) by the fraction of red galaxies, $f_{\rm red}(z)$, as a function of redshift. The physical interpretation of the intrinsic alignment contribution to this cross-correlation is that the same density field that acts to align galaxies is lensing the CMB. The angular power spectrum corresponding to this component is given by

\begin{equation}
C^{I\kappa_{\rm CMB}}(l) = \int_{0}^{\infty} dz \frac{H(z)}{c} \frac{W^{\rm CMB}(z)f_{\rm red}(z)p(z)}{\chi^2(z)} P_{\delta I}(l=k\chi,z),
\label{eq:Cl_iacmb}
\end{equation}

\noindent where $p(z)$ is the redshift distribution of galaxies with shapes (normalized to an integral of $1$ over the full redshift range probed), $\chi$ is the comoving distance as a function of $z$ and $W^{\rm CMB}(z)$ is the CMB lensing kernel given by \citep{Lewis06}

\begin{equation}
W^{\rm CMB}(z) = \frac{3\Omega_mH_0^2}{2c}\frac{(1+z)}{H(z)}\chi(z)\left(1-\frac{\chi(z)}{\chi^\star}\right),
\end{equation}
where $\chi^\star$ is the comoving distance to the surface of last scattering.
Note that Eq. (\ref{eq:Cl_iacmb}) does not depend on the bias of the galaxy sample with respect to the overdensity field. In comparison, typical measurements of the intrinsic alignment strength from overdensity-shape cross-correlations rely on simultaneously constraining the galaxy bias from clustering measurements.

The cross-correlation of CMB lensing with galaxy lensing cannot be measured in isolation. Its power spectrum is given by

\begin{equation}
C^{G\kappa_{\rm CMB}}(l) = \int_0^{\infty} dz \frac{H(z)}{c} \frac{W^G(z) W^{\rm CMB}(z)}{\chi^2(z)}P(l=k\chi(z),z),
\label{eq:Cl_gcmb}
\end{equation}

\noindent where the galaxy lensing kernel is 

\begin{equation}
W^G (z) = \frac{3\Omega_mH_0^2}{2c}\frac{(1+z)}{H(z)}\chi(z) \int_z^\infty dz' p(z') \frac{\chi(z')-\chi(z)}{\chi(z')}.
\end{equation}

The cross-correlation of galaxy shapes and CMB lensing, $C^{\kappa_S\kappa_{\rm CMB}}$, is the sum of lensing and alignment contributions. Alignments and lensing distort the ellipticities with opposite signs: alignments tend to decrease the lensing amplitude, and thus the two signals need to be modeled jointly. 

The associated uncertainty in measuring this cross-correlation, which combines accounts for the constributions of cosmic variance, galaxy shape noise and noise in the lensing maps is given by

\begin{equation}
  {\rm Var} [C^{(G+I)\kappa_{\rm CMB}}(l)] = [(2l+1)f_{\rm sky}]^{-1}\left[\left(C_l^{GG}+C_l^{GI}+C_l^{II}+\frac{\sigma_\gamma^2}{n_{\rm \Omega}}\right)\left(C_l^{\kappa_{\rm CMB}\kappa_{\rm CMB}}+N_l^{\kappa_{CMB}}\right)+\left(C_l^{(G+I)\kappa_{\rm CMB}}\right)^2\right],
\label{eq:var}
\end{equation}
where $n_{\Omega}$ is the effective surface number density of galaxies with shapes per steradian \citep{Chang13} and $\sigma_\gamma$ is the shape noise per component of the ellipticity. 
The $S/N$ for the detection of the cross correlation of CMB lensing and galaxy shapes is given by
\begin{equation}
\left(\frac{S}{N}\right)^2 = \sum_{l=l_{\rm min}}^{l_{\rm max}}\frac{(C^{(G+I)\kappa_{\rm CMB}}(l))^2}{{\rm Var} [C^{(G+I)\kappa_{\rm CMB}}(l)]}.
\label{eq:sn}
\end{equation}

To quantify the signal-to-noise ratio for future surveys we must compute the CMB lensing reconstruction noise, $N_l^{\kappa_{\rm CMB}}$. We use the quadratic-estimator formalism of \cite{Hu02} to derive the noise level appropriate for an ACT-like survey \citep{Das14}. For a current generation survey such as ACTPol \citep{vanEngelen14}, the quadratic estimator predicts that the improved map sensitivity and the inclusion of polarisation measurements results in a decrease by a factor of $2$ of the noise level with respect to the ACT experiment. Similarly, we find a factor of $3$ reduction for the next generation AdvancedACT survey \citep{Calabrese14}. Note that we assume a white-noise-only primary CMB noise spectra, neglecting atmospheric noise; this slightly underestimates the true lensing reconstruction noise. This effect will be less important for lensing reconstruction from polarisation. The assumed noise for the galaxy shear convergence field is also optimistic, as we only consider pure shape noise and cosmic variance, and we do not model the convergence field reconstruction.

\section{Modeling the early-type galaxy population}
\label{sec:redgal}

For a generic magnitude-limited survey in $r$ or $i$ optical bands, we construct in this section a model for the fraction of red galaxies subject to alignment. We follow \cite{Joachimi11} in modeling the red (early-type) galaxy population, in conjunction with the observed red galaxy Schechter luminosity function \citep{Schechter76} observed by \cite{Faber07}. The former measure the luminosity function of red galaxies in the combination of SDSS \citep{Bell03} and 2dF \citep{Madgwick02} surveys at low redshift ($z<0.1$), and COMBO-17 \citep{Bell04} and DEEP2 \citep{Willmer06} surveys at high redshift ($z>0.2$). We fit these measurements in the range $0.04<z<1.2$ and we extrapolate to $z<0.04$ or $z>1.2$ based on this linear fit. The luminosity function of red galaxies in \cite{Faber07} is measured in $B$ band (Vega). We apply the relevant conversions to AB \citep{Willmer06} to relate the luminosity function to the observed apparent magnitudes of galaxies in CS82. 

\cite{Joachimi11} made a forecast of the intrinsic alignment angular power spectrum in an $r$-band magnitude limited survey. Here, we find it necessary to work also with an $i$-band limited survey, which resembles more closely the CS82 survey. Given a limiting magnitude $r_{\rm lim}$, the minimum $B$-band absolute magnitude of the galaxies observed at a given redshift is given by 
\begin{equation}
  M_{\rm min}(z,r_{\rm lim}) = r_{\rm lim} - \left[ 5\log_{10}\left( \frac{D_L}{{\rm Mpc}}\right) +25 +k_{{\rm red},r}(z)\right]+(B-r)(z),
  \label{eq:mmin}
\end{equation}
where $k_{{\rm red},r}$ is the $K$-correction for red galaxies in the $r$ band, $D_L$ is the luminosity distance in Mpc. An analogous expression holds for an $i$-band limited survey. We approximate the evolution of the rest-frame color as $(B-r)(z)\simeq (g-r)(z)$, since $g$ and $B$ cover the same wavelength range, and similarly for $B-i$. \citet{Joachimi11} assumed that $B-r$ did not evolve significantly (compared to the uncertainties in choice of luminosity function), but this ceases to be true beyond $z\sim 1$, so we account for evolution explicitly in the last term of Eq. (\ref{eq:mmin}). This redshift evolution is not applied in the case of the reference magnitude, $M_0(z)$, since Eq. \ref{eq:AIJ11} is fit to the data under the assumption of no evolution of $B-r$. 

$K+e$ corrections account for the intrinsic evolution of galaxies and for the fact that a given filter overlaps with different parts of the spectral energy distribution of a galaxy at different redshifts. These corrections are needed to go from absolute to apparent magnitudes. To calculate $K+e$ corrections, we assume that red galaxies can be modeled as a single burst of star formation at $t=13.32$ Gyr (``Passive'' model). Alternatively, we consider a single burst at the same redshift that is followed by a constant star formation rate that accounts for $5\%$ of the final mass (``Passive+SF'' model). This is the same procedure as adopted by \cite{Wake06}, who find that these two models reproduce the $g-i$ colours of LRGs. We have verified there is good agreement between the observed LRG spectrum observed by \cite{Wake06} and the model templates constructed from \cite{Bruzual03}. We have also found good agreement between for the color evolution ($g-r$ and $r-i$) computed by \cite{Wake06} and for $g-i$ and $r$-band $K+e$ corrections as a function of redshift \citep{Joachimi11} for the two models. \cite{Wake06} find a small offset between the $g-r$ and $r-i$ colours of LRGs and in the models. For the purpose of this work, we rely on $g-i$ colours, which are well reproduced by the models.

The minimum absolute magnitude corresponds to a minimum luminosity,
\begin{equation}
\frac{L_{\rm min}(z,r_{\rm lim})}{L_0(z)} = 10^{-0.4[M_{\rm min}(z,r_{\rm lim})-M_0(z)]}.
\end{equation}
The observed number density of red galaxies per comoving volume is obtained by integrating the Schechter $B$-band luminosity function for red galaxies above the limiting magnitude, 
\begin{equation}
n_{V,{\rm red}}(z) = \phi^\star(z) \Gamma\left(\alpha+1,\frac{L_{\rm min}(z,r_{\rm lim})}{L^\star(z)} \right)
\label{eq:nred}
\end{equation} 
and the fraction of red galaxies is obtained as the ratio between Eq. (\ref{eq:nred}) and the total comoving number density of galaxies with measured ellipticities in the survey,
\begin{equation}
f_{\rm red}(z) \equiv \frac{n_{V,{\rm red}}(z,r_{\rm lim})}{n_{V,{\rm tot}}(z)} = \frac{\chi^2(z)\chi'(z)}{n_\Omega p(z)} \int_{L_{\rm min}(z,r_{\rm lim})}^\infty dL \phi(L,z)
\label{eq:redfrac}
\end{equation}
Notice that the use of the effective number of galaxies in the survey in the Eq. (\ref{eq:redfrac}) implicitly assumes that all galaxies observed by \cite{Faber07} to construct the luminosity function pass shape cuts; this is a conservative assumption. If fewer red galaxies had measured shapes, the alignment contamination would decrease. For example, \cite{Joachimi11} found that only $50\%$ of the galaxies in the MegaZ-LRG sample had measured shapes. For modeling the alignments of blue galaxies, we assume that the fraction of blue galaxies is given by $f_{\rm blue}(z)=1-f_{\rm red}(z)$.

Finally, for modeling the luminosity dependence of the strength of alignment (Eq. \ref{eq:AIJ11}), we compute the mean $r$-band luminosity of the galaxies to some power $\beta$,

\begin{equation}
\frac{\langle L^\beta\rangle}{L_0^\beta} = \left[\frac{L^\star(z)}{L_0(z)} \right]^\beta \frac{\Gamma\left(\alpha+\beta+1,L_{\rm min}/L^\star\right)}{\Gamma\left(\alpha+1,L_{\rm min}/L^\star\right)}
\end{equation}
with $\beta=1.13$ for the \cite{Joachimi11} alignment model constraints.


\section{Results}
\label{sec:results}

We now apply the model of the number of early-type galaxies developed in Section \ref{sec:redgal} to compute the intrinsic alignment contamination to the cross-correlation of weak gravitational lensing and CMB lensing from the CS82 survey and the ACT CMB experiment as measured by \citet{Hand13}. We assume $p(z)$ as given in \citet{Hand13}, 
\begin{equation}
  p(z) \propto \frac{z^a+z^{ab}}{z^b+c}
  \label{eq:pofz}
\end{equation}
and we adopt their best fit parameters for $\{a,b,c\}$. We normalize the $p(z)$ to $1$ in the range $0<z<6$. We also adopt as fiducial properties of the CS82 survey: a limiting magnitude of $i_{\rm lim}=24.1$, a shear dispersion per component of $\sigma_\gamma=0.26$, $n_\Omega=8.5$ effective weighted galaxies with shapes per arcmin$^2$. The CS82 sample of galaxies with shapes includes some galaxies above $i_{\rm lim}=24.1$. For this reason we also repeat our calculations assuming a deeper survey, to $i_{\rm lim} = 24.7$.

\begin{figure}
\centering
\includegraphics[width=0.4\textwidth]{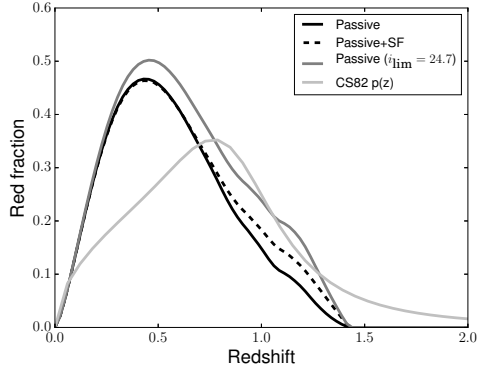}
\caption{Estimated fraction of red galaxies observed in the CS82 imaging survey as a function of redshift. The fraction is obtained assuming that the survey is limited in $i$-band to an apparent magnitude of $24.1$. The solid lines correspond to the case where red galaxies evolve passively from $t=13.32$ Gyr to the present, while the dashed line corresponds to galaxies with a burst of star formation at $t=13.32$ Gyr and constant star formation until $z=0$, which contributes to form $5\%$ of the final mass. The dark gray line corresponds to a deeper survey, with $i_{\rm lim}=24.7$. The light gray line represents the redshift distribution of galaxies in the CS82 survey, arbitrarily normalized.}
\label{fig:redfrac}
\end{figure}

In Figure \ref{fig:redfrac} we show the estimated fraction of early-type galaxies as a function of redshift in the CS82 survey. The black solid line represents the model in which all galaxies evolve passively from $t=13.32$ Gyr, when they formed all their stars, to the present epoch. The dashed line corresponds to the model where red galaxies form $95\%$ of their stars at a light travel time $t=13.32$ Gyr and continue forming stars at a constant rate until today. The red fraction decreases at high redshifts, which is consistent with the results of \cite{Faber07}, who find that the comoving number density of blue galaxies is roughly constant to $z\sim 1$ while that of red galaxies monotonically increases towards $z=0$. At $z\lesssim 0.4$, the fraction of red galaxies also decreases, but this behaviour is quite sensitive to the redshift dependence of $dN/dz$ in the CS82 survey and can change quite significantly within the allowed range of uncertainty for that function. For $i_{\rm lim}=24.1$, there are $2.1$ red galaxies per arcmin$^2$. This corresponds to $\simeq 26\%$ of the effective number of galaxies per arcmin$^2$ in CS82.

\begin{figure}
\centering
\includegraphics[width=0.33\textwidth]{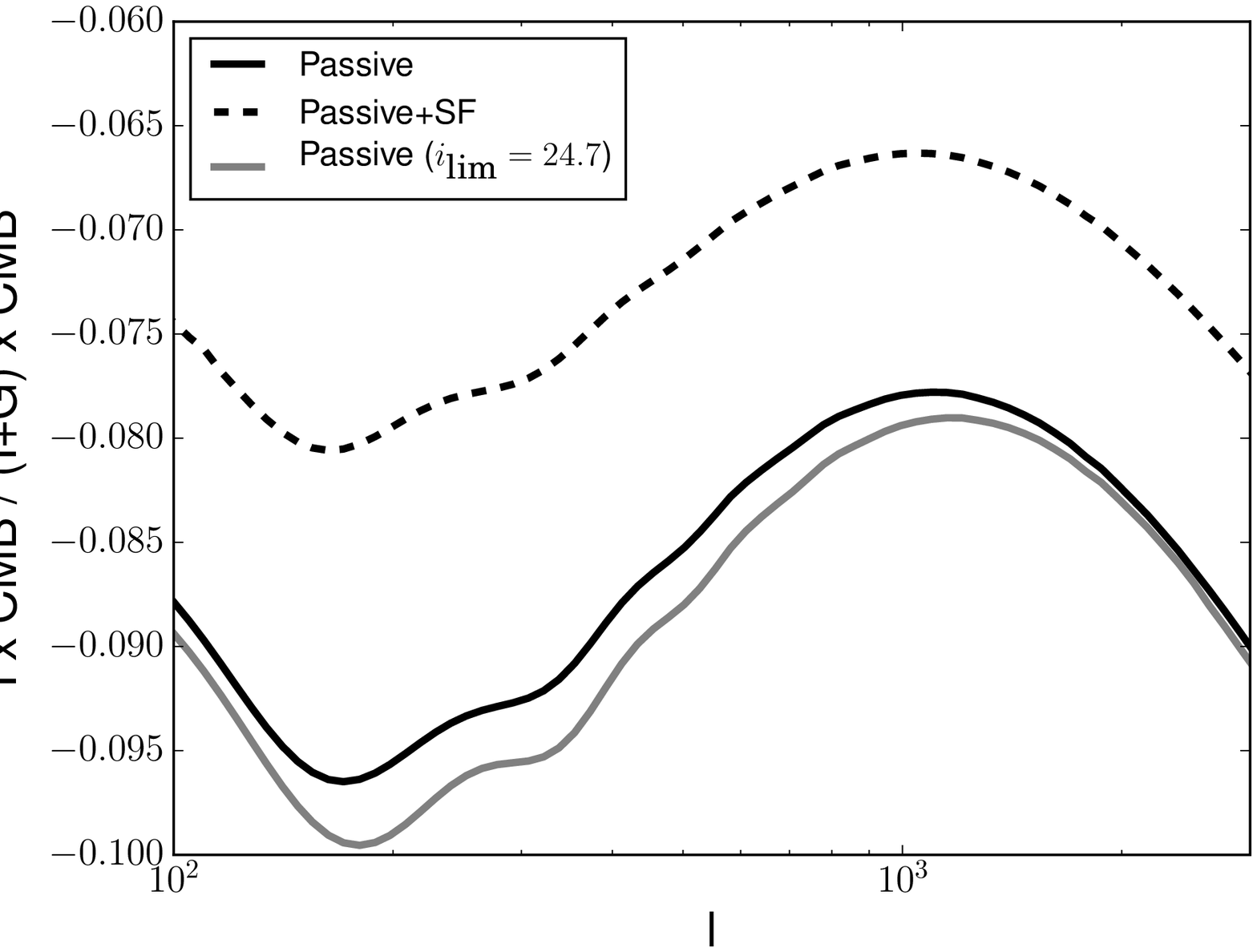}
\includegraphics[width=0.33\textwidth]{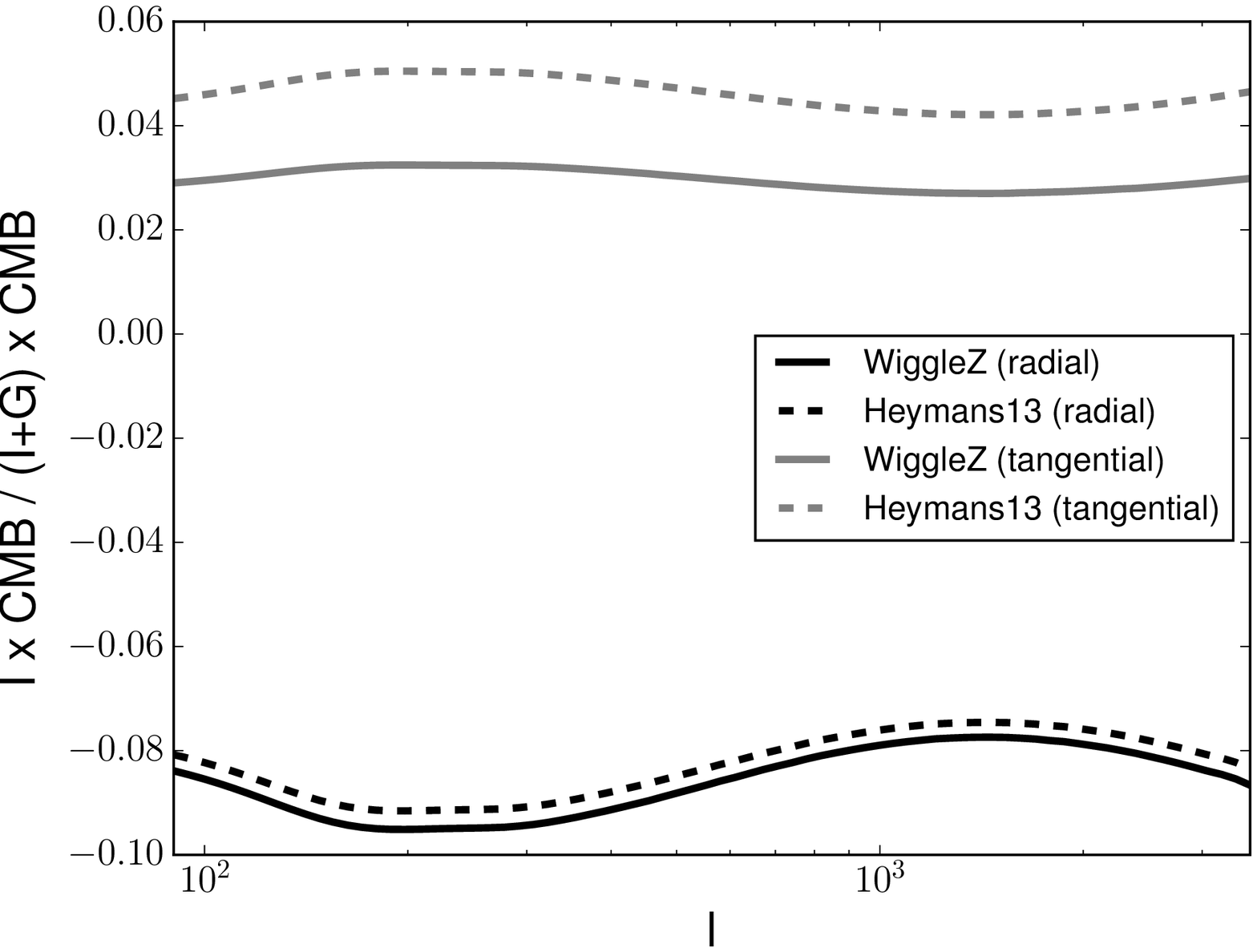}
\includegraphics[width=0.33\textwidth]{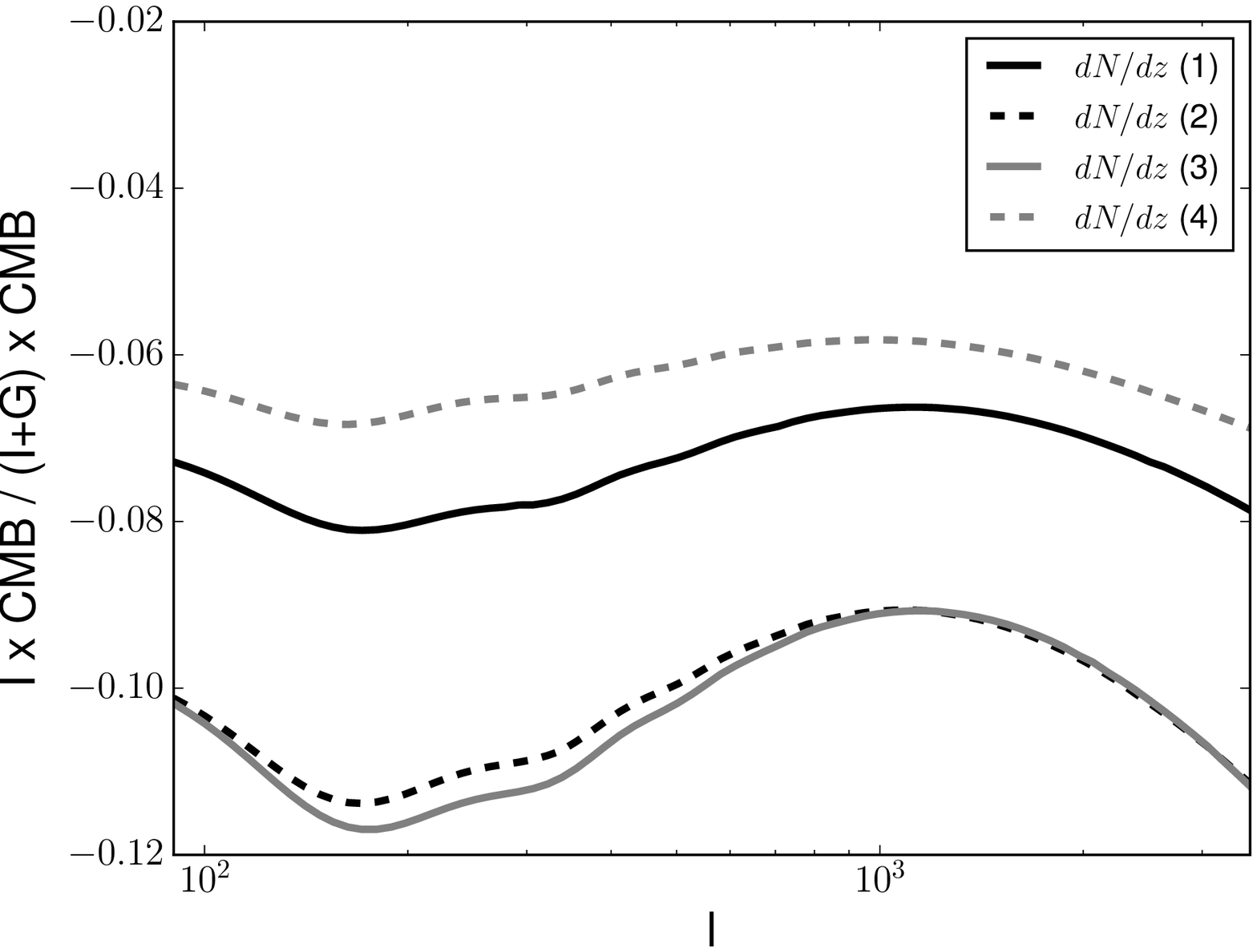}
\caption{The ratio between intrinsic alignments-CMB lensing cross-correlation and weak lensing convergence-CMB lensing cross-correlation for CS82. The left panel represents our fiducial scenario, where only red galaxies up to a redshift of $\sim 1.4$ contribute to alignments. The solid lines are obtained by assuming that red galaxies evolve without forming stars from $t=13.32$ Gyr to the present, while the dashed line corresponds to galaxies with a burst of star formation at $t=13.32$ Gyr and constant star formation until $z=0$, which contributes $5\%$ of the mass at $z=0$. The gray line corresponds to a deeper survey, with $i_{\rm lim}=24.7$, to account for the faint tail of the CS82 sources. The middle panel shows the constraints on contamination from blue galaxy alignments from \citet[$95\%$ confidence level]{WiggleZ} and \citet[$1\sigma$]{Heymans13} within the current uncertainty levels. Notice that a small amount of tangential alignments are allowed within the current error bars. The right panel shows the impact of the uncertainties in the $dN/dz$ for CS82 applying the different functional forms suggested by \citet{Hand13}.
The effect of shifting the position of the peak of $dN/dz$ to higher (lower) redshift decreases (increases) the passive galaxy alignment contamination to $8.1\%$ 
($11\%$) 
at its maximum ($l\simeq 200$) and for the passive galaxy model. These scenarios correspond to labels (1) and (2), respectively. Increasing (decreasing) the predominance of the redshift tail changes the alignment contamination to up to $12\%$ 
(down to $6.9\%$) 
at $l\simeq200$; curves (3) and (4), respectively.}
\label{fig:fraction}
\end{figure}

\begin{figure}
\centering
\includegraphics[width=0.5\textwidth]{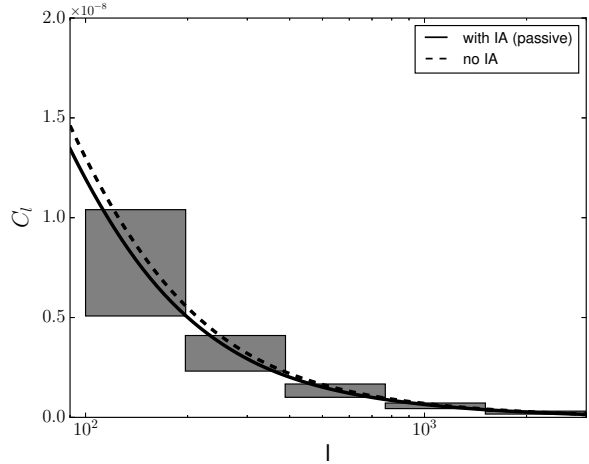}
\caption{The CMB lensing-galaxy shape cross-correlation, as forecast from our model for CS82$\times$ACT as in \citet{Hand13}. The dashed line represents the case with no intrinsic alignments, while the solid line takes them into account, lowering the amplitude of the cross-correlation by $\approx 10\%$. The intrinsic alignment impact is well within the error bars of the measurement, which are shown centered on the model of galaxy shapes that includes intrinsic alignments. The forecast signal-to-noise on $C_l^{\kappa_{\rm CMB}G}$ is high compared to the actual measurement of \citet{Hand13}. This can be attributed to neglecting the atmospheric noise contribution in the ACT temperature map which leads to an underestimation of $N_l^{\kappa_{\rm CMB}}$, and also to a subestimation of the noise in the galaxy shear convergence, which we model as a combination of shape noise and cosmic variance.} 
\label{fig:full}
\end{figure}

In Figure \ref{fig:fraction}, we show the fractional contribution of intrinsic alignments to the angular power spectrum $C_l^{(G+I)\kappa_{\rm CMB}}$. In the left panel, the lines correspond to the same models as in Fig. \ref{fig:redfrac}. Intrinsic alignments contaminate the total galaxy shape-CMB lensing convergence angular power spectrum at a level of $<10\%$ in the CS82 and ACT cross-correlation measurement and for the passive evolution model. This case is also shown in Figure \ref{fig:full}, where we show the angular power spectrum of the CMB lensing-galaxy lensing cross-correlation for the combination of ACT and CS$82$ in the cases with and without intrinsic alignments. The impact of intrinsic alignments is clearly within the current error bars. Moreover, our noise estimates are optimistic both for the CMB lensing convergence and of the galaxy convergence field, as discussed in Section \ref{sec:theory}.

The uncertainty on the star-formation activity of LRGs results in about $1.6\%$ decrease of this estimate, and uncertainty in magnitude limit of the survey, an increase of $0.3\%$.
Due to the paucity of red galaxies at redshift $\geq 1.2$, this result is not too sensitive to the details of the luminosity function of red galaxies at those redshifts. The contribution of red galaxies with $z>1.2$ to the galaxy lensing-CMB lensing cross-correlation is only $0.9\%$. 
Our results do not change significantly if we consider the possibility that the luminosity function of red galaxies remains constant above $z=1.2$. 
Allowing the intrinsic alignment amplitude to increase within the $1\sigma$ constraint obtained by \citet{Joachimi11}, the maximum contamination from alignments increases to $10.6\%$.

We consider the contribution of late-type alignments by using the non-linear alignment model with the blue galaxy alignment strength constrained by \citet{WiggleZ} and a blue fraction given by $1-f_{\rm red}$. This amplitude for late-type alignments, while consistent with null, could lead to a $9.5\%$ contamination within the $95\%$ confidence level constraints on the alignment strength of blue galaxies. Assuming that we are allowed to extrapolate from low and intermediate redshifts ($z\sim 0.6$) to the full redshift range considered for CS82,  \citet{WiggleZ} constraints on blue galaxy alignments are the most stringent to this date. If we consider the $1\sigma$ inferred late-type alignment amplitude of \citet{Heymans13}, which probes up to $z=1.3$, we obtain a maximum contamination of $9.2\%$. This estimate includes the contribution of blue early-types \citep{Schawinski09} up to $z=1.3$, which would have been mis-classified as late-types based on their spectral energy distribution. Blue galaxy alignments contamination is shown in the middle panel of Figure \ref{fig:fraction}. Contamination from tangential alignments (positive bounds) is allowed at $1\sigma$ from \citet{Heymans13} and within the $95\%$ confidence levels from \citet{WiggleZ}.

Above $z=1.3$, there remains the possibility that the majority of early-types are blue and subject to alignments. Current observations cannot rule out this possibility. As a worst case scenario, we assume that {\it all} galaxies at $z>1.2$ are subject to alignments. For a magnitude limited survey, red galaxies observed at higher redshifts will be on average brighter than the average of the population at lower redshift. Because the alignment strength increases with luminosity, the observed early-type galaxy population at higher redshifts has an enhanced $A_I$ compared to galaxies at lower redshifts. This effect competes with the paucity of early-type galaxies at $z\gtrsim 1$. This is a Malmquist bias-like effect, and it would not be present in a volume-limited survey. Malmquist bias acting on the high redshift galaxies can lead to arbitrarily large alignment amplitudes in the current model (see Eq. \ref{eq:AIJ11}), and hence we neglect the luminosity dependence of $A_I$ in this case. Instead, we match the high redshift alignment strength to the maximum low redshift value for this estimate ($A_I=16.09$, \citealt{Joachimi11}). In this case, the alignment contribution to the galaxy shape-CMB lensing cross-correlation can be as high as $\sim 60\%$ of the total signal.

This suggests the cross-correlation with CMB lensing is a promising avenue to constrain the amplitude of alignments at high redshift, as originally suggested by \citet{Hall14} and \citet{Troxel14a}. Indeed, \citet{Kitching14} have suggested that intrinsic alignments in the KiDS survey could be calibrated through the use of CMB lensing information from ACTPol. We have made use of the shear-CMB lensing correlation of \citet{Hand13} to constrain the allowed contribution from intrinsic alignments using our fiducial model and our fiducial cosmology. If alignments were the only systematics at play, a best-fit $\sim 24\%$ contamination would be required to bring the measurements in agreement with the $\Lambda$CDM expectations. Changing the assumed $dN/dz$ for CS82 as in \citet{Hand13}, this result can vary from $15\%$ to $30\%$. For the fiducial redshift distribution, a $60\%$ contamination would be allowed at the $68\%$ confidence level, and hence our worst case scenario cannot be fully rejected.
This result also suggests that our lack of knowledge of intrinsic alignments at high redshifts can hamper CMB lensing-galaxy lensing cross-correlations and this could be mitigated by removing the high redshift tail of lensed galaxies from the cross-correlation, thus restricting to the redshift range where alignments are well characterized by current observations. Removing the high redshift tail of the CS82 galaxy sample would result in $30\%$ reduction in the signal-to-noise of the measurement; but in practice the feasibility of this exercise would be dependent on the quality of the photometric redshifts and the ability to separate the high-$z$ tail.


The redshift distribution of galaxies in CS82 is subject to some uncertainty. \citet{Hand13} attributed their original discrepancy between the observed amplitude of the cross-correlation of galaxy lensing and CMB lensing and the expected value based on {\it Planck} cosmology \citep{Planck} to uncertainties in the $p(z)$, which can result in $10-20\%$ changes in the amplitude of the expected cross-correlation. They consider the effects of changing the $p(z)$ by changing the peak value of the distribution in $\Delta z = \pm 0.1$, and they also consider modifying the parameter $b$ in Eq. (\ref{eq:pofz}) by $\pm 30\%$ to change the relative weight of the high redshift tail. 
We have considered similar modifications to the $p(z)$ and observed that shifting the peak of the distribution to lower (higher) redshift increases (decreases) the passive galaxy alignment contamination to
$\simeq 11\%$ 
($8.1\%$) 
at its maximum ($l\simeq 200$) and for the passive galaxy model. The effect of increasing (decreasing) the predominance of the redshift tail changes the alignment contamination by up to 
$12\%$ 
(down to $6.9\%$)  
at $l\simeq200$. The different $dN/dz$ cases are shown in the right panel of Fig. \ref{fig:fraction}.

Previous work by \cite{Hall14}and \cite{Troxel14a} estimated a contamination from intrinsic alignments to galaxy lensing-CMB lensing cross-correlation of $15\%$ using low redshift ($z\sim 0.1$) constraints on the contribution from intrinsic alignments. In this work, we have used up-to-date measurements of intrinsic alignments at higher redshifts (in better agreement with the CS82 $dN/dz$) and as a function of redshift, galaxy color and absolute magnitude to obtain new estimates of contamination to galaxy lensing-CMB lensing cross-correlation. Our fiducial model indicates that the contribution of red galaxies is $\lesssim 10\%$ under pessimistic assumptions. The blue galaxy alignments contribution is consistent with null, but it could have a comparable contribution to that of early-type galaxies ($\simeq 9.5\%$) within the current uncertainty level. Overall, the fiducial results are in good agreement with \cite{Hall14} and \cite{Troxel14a} estimates. The contribution of intrinsic alignments remains non-negligible and should be taken into account in this type of cross-correlation measurement. Moreover, large uncertainties remain regarding the potential contribution of alignments at higher redshifts. 

\citet{Liu15} found a discrepancy between the expected CMB lensing-galaxy lensing cross-correlation from {\it Planck} and CFHTLenS, at much higher significance than the \citet{Hand13} measurement. The redshift distribution and magnitude limit of the CFHTLenS shear sample are sufficiently similar to CS82 that we do not expect major differences in the estimate of alignment contamination. Assuming no other systematics are at play, it would be challenging to come up with an alignment model that could reconcile at the same time the \citet{Liu15} and \citet{Hand13} results with the expectations from {\it Planck} and WMAP9. It is likely that other systematics are at play.

\section{Limitations for future surveys}
\label{sec:forecast}

Future surveys will detect the cross-correlation of CMB lensing and galaxy lensing to high significance. For simplicity, we neglect the impact of intrinsic alignments to estimate the signal-to-noise of this measurement for a next generation of experiments. 
We assume that $n_{\Omega}$ increases to $26$ arcmin$^{-2}$ as expected for the {\it LSST} survey, with a redshift distribution of 
\beeq
\frac{dN}{dz}\propto z^\alpha \exp{\left(\frac{z}{z_0}\right)^\beta},
\eneq
where $\alpha=1.24$, $z_0=0.51$, $\beta=1.01$, and spanning a range of $0.1<z<3$ \citep{Chang13}. The median redshift of LSST galaxies is $z=0.83$ \citep{LSST}. For the noise in the CMB lensing reconstruction, we first assume a decrease by a factor of $3$ for the completed ACTPol experiment \citep{vanEngelen14} relative to the ACT experiment \citep{Das14}. {\it LSST} will cover $18,000$ deg$^2$ in a ``deep-fast-wide" mode, hence the cross-correlation will rather be limited to the area of overlap with ACTPol. ACTPol is expected to cover $3000$ deg$^2$ \citep{Niemack10} and we thus assume that $f_{\rm sky}$ increases to $0.07$ in comparison to $0.003$ for the ACT-CS82 cross-correlation. With {\it LSST} and ACTPol, the cross-correlation of galaxy lensing and CMB lensing could be detected with a signal-to-noise ratio $\simeq 10$ times larger than current measurements. 
We also consider an AdvACT-like survey \citep{Calabrese14}, with a full overlap with LSST. AdvACT will take data from 2016 to 2018, and is thus in good synergy with LSST. For AdvACT, we expect the signal-to-noise for detection of CMB lensing-galaxy lensing cross-correlation will be $\simeq 30$  times larger than for current measurements. For both experiments, the improvement of the $S/N$ over the combination of CS$82$ and ACT is mostly driven by the increase in the area. However, we emphasize that $S/N$ forecast tend to be optimistic given the current modelling of the CMB lensing noise and the noise of the galaxy convergence, as discussed in Section \ref{sec:results}.

The limitations of the applicability of our formalism to future surveys are the following. First, intrinsic alignments of red galaxies have been so far constrained up to $z=1.3$ \citep{Heymans13}, below the maximum redshift probed by future surveys. Several uncertainties remain at higher redshift. It is unclear whether the distinction between red and blue galaxies can separate the populations with and without intrinsic alignments at high redshift. The alignment mechanism most likely depends on the galaxy dynamics and formation history, rather than color. The fraction and $A_I$ of aligned galaxies at higher redshift remains unconstrained and we have shown that galaxies at high redshift could contribute very significantly to the cross-correlation of CMB lensing and shear ($\simeq 50\%$ of the total signal). Ideally, constraints on alignment strength would be done jointly with luminosity function estimates, which would improve the modeling of $A_I$ and of the aligned fraction.

Moreover, while studies of the evolution of red galaxy alignments \citep{Joachimi11,Singh14} have shown it to be consistent with the tidal alignment model \citep{Catelan01,Hirata04}, at higher redshifts future studies must test the assumption that galaxies obtain their alignments at formation. Even under the optimistic assumption that current models can be extrapolated to $z>1.2$, the uncertainty in the alignments of blue galaxies is still so significant as to contribute up to a similar percentage of contamination as early-type galaxies to the shape-CMB lensing correlation of ACT and CS82.

Third, the power-law dependence of $A_I$ on luminosity has been measured from magnitude-limited surveys which do not probe as faint galaxies as future surveys will probe for cosmic shear measurements. As future surveys reach fainter intrinsic luminosities, it is possible that the power-law exponent of Eq. (\ref{eq:AIJ11}) will change. For example, \citet{Tenneti14b} find a shallower luminosity scaling than \citet{Joachimi11} for intrinsic alignments in hydrodynamical numerical simulations.

The scale-dependence of the intrinsic alignment signal is currently poorly understood on small scales. The non-linear alignment model is based on assuming that the physics is linear, and that the linear power spectrum can be simply replaced by its non-linear analogue in Eq. (\ref{eq:iaps}) \citep{Bridle07}. A halo model has been developed to provide additional degrees of freedom on the description of alignments on small scales \citep{Schneider10}, but it does not fully capture the effects of the large-scale structure on alignments, i.e., the effect of filaments \citep{Altay06}. Recently, the validity of these models has been tested with observations of alignments in the LOWZ galaxy sample by \cite{Singh14}. These authors find that the combination of the halo model on small scales ($<1$Mpc$/h$) with the linear treatment on large scales does not reproduce the observed scale-dependence of the alignment signal, and hence a combination of the one-halo term and the non-linear alignment model is typically preferred. Nevertheless, this combination is hard to interpret physically. The recent work by \citet{Blazek15} marks an improvement in this direction, building a more thorough morel for alignments on small scales. These uncertainties in the modeling of alignments at intermediate scales can propagate into cosmological parameter biases in future surveys. 

Finally, the redshift distribution and selection effects on galaxies with shapes are subject to significant uncertainties. We showed in the previous section that the uncertainty in the CS82 $dN/dz$ has a significant impact on the estimate of the contamination from intrinsic alignments.

In preparation for large-scale imaging surveys such as {\it Euclid} and {\it LSST}, these issues need to be addressed. While deeper surveys would provide a wealth of information on the scale-dependence, luminosity dependence and redshift dependence of alignments, other promising methods are starting to be developed. Hydrodynamical cosmological simulations are beginning to explore open questions in the physics of intrinsic alignments \citep{Tenneti14a,Tenneti14b,Codis14}. Puzzlingly, simulations suggest that blue galaxy alignments, consistent with null in low redshift observations \citep{WiggleZ,Heymans13}, could contribute significantly to the alignment signal at $z\gtrsim 1$ \citep{Codis14}. On the other hand, \cite{Tenneti14b} reproduce the observed alignments of red galaxies at low redshifts, and are able to make predictions for the amplitude of alignments for typical {\it LSST} source galaxies.

\section{Conclusions}

The cross-correlation of galaxy lensing and CMB lensing, first measured by \citet{Hand13}, probes the dark matter distribution at intermediate redshifts and can help determine multiplicative biases in shear measurements \citep{Vallinotto12,Das13}. This cross-correlation is, however, not measured in isolation - intrinsic alignments contribute an additional component.

In this work, we have applied a model of the red galaxy population based on \citet{Joachimi11} to estimate the intrinsic alignment contamination to the cross-correlation of the convergence field derived for galaxy shapes and the lensing of the CMB. In the case of the CS82 imaging survey, combined with CMB lensing measurements from ACT, we have estimated the contribution of intrinsic alignments to be $9.6\%$. Previous works had placed an upper limit for this contamination at $15\%$ \citep{Troxel14a,Hall14} relying on low redshift estimates of the alignment amplitude \citep{Brown02}. Using more up-to-date observational constraints of alignments that include colour, redshift and absolute magnitude dependence, our fiducial results suggest that those previous estimates are not significantly modified. Apart from the contribution of early-type galaxies, blue galaxy alignments could contribute to the shear-CMB lensing cross-correlation by up to a similar amount within current uncertainties.

However, at $z>1.2$, alignments remain unconstrained. If all galaxies above this redshift are subject to alignments with a similar strength as low redshift LRGs, the cross-correlation of CMB lensing and galaxy shear could be largely decreased (with alignments contributing up to $\sim 60\%$ of the total signal). This large uncertainty suggests that this cross-correlation could be used to estimate alignments at high redshifts, as initally suggested by \citet{Hall14} and \citet{Troxel14a}. It also suggests that, for current CMB lensing-galaxy lensing cross-correlations, removing sources in the high redshift tails would be a viable option for reducing the uncertainties due to intrinsic alignments.

In preparation for future surveys \citep{Laureijs11,DESCWhite}, the scientific community is currently planning to achieve better constraints on the redshift dependence, luminosity dependence and the galaxy population subject to alignments, as well as the evolution of the luminosity function of this population using on-going surveys such as Hyper Suprime-Cam\footnote{\url{http://www.naoj.org/Projects/HSC/}}, the Dark Energy Survey\footnote{\url{http://www.darkenergysurvey.org}}, the Kilo-Degree Survey\footnote{\url{http://kids.strw.leidenuniv.nl}} and Pan-STARRS\footnote{\url{http://pan-starrs.ifa.hawaii.edu/public/}}.

\section*{Acknowledgments}
NEC is supported by a Beecroft Postdoctoral Research Fellowship. JD and RA are supported by ERC grant 259505.
We thank Blake Sherwin, Alexie Leauthaud, Nick Hand, the CS82 and ACT teams for useful exchanges about their work on galaxy lensing-CMB lensing cross-correlation \citep{Hand13} and for sharing their data with us. We are grateful to Rachel Mandelbaum for comments that helped improve this work.

\bibliographystyle{mn2e}
\bibliography{cmbia}

\label{lastpage}

\end{document}